\documentclass[pdflatex,sn-nature]{sn-jnl}

\usepackage{graphicx}%
\usepackage{multirow}%
\usepackage{amsmath,amssymb,amsfonts}%
\usepackage{amsthm}%
\usepackage{mathrsfs}%
\usepackage[title]{appendix}%
\usepackage{xcolor}%
\usepackage{textcomp}%
\usepackage{manyfoot}%
\usepackage{booktabs}%
\usepackage{algorithm}%
\usepackage{algorithmicx}%
\usepackage{algpseudocode}%
\usepackage{listings}%

\begin{document}

\title[Mutually phase-stable tunable attosecond soft 
X-ray pulse pairs from a free-electron laser]{Mutually phase-stable tunable attosecond soft 
X-ray pulse pairs from a free-electron laser}


\author*[1,2,3]{\fnm{River}\sur{Robles}}\email{riverr@stanford.edu}

\author[1,2,4]{\fnm{Kurtis}\sur{Borne}}

\author[1,2]{\fnm{Mathew}\sur{Britton}}

\author[1]{\fnm{David}\sur{Cesar}}

\author[1,2,3]{\fnm{Paris}\sur{Franz}}

\author[1,2,3]{\fnm{Veronica}\sur{Guo}}

\author[1]{\fnm{Gabriel}\sur{Just}}

\author[1,2]{\fnm{Kirk A.}\sur{Larsen}}

\author[1]{\fnm{Randy}\sur{Lemons}}

\author[1,2]{\fnm{Vyacheslav}\sur{Leshchenko}}

\author[1,2]{\fnm{Xiang}\sur{Li}}

\author[1,2]{\fnm{Ming-Fu}\sur{Lin}}

\author[1,2]{\fnm{Razib}\sur{Obaid}}

\author[1]{\fnm{Nicholas}\sur{Sudar}}

\author[1,2,3]{\fnm{Jun}\sur{Wang}}

\author[1]{\fnm{Zhen}\sur{Zhang}}

\author*[1,2]{\fnm{James}\sur{Cryan}}\email{jcryan@slac.stanford.edu}

\author[1,2]{\fnm{Taran}\sur{Driver}\email{tdd14@slac.stanford.edu}}

\author*[1,2]{\fnm{Agostino}\sur{Marinelli}}\email{marinelli@slac.stanford.edu}

\affil[1]{\orgname{SLAC National Accelerator Laboratory}, \orgaddress{\city{Menlo Park}, \state{California} \postcode{94025}, \country{USA}}}

\affil[2]{\orgdiv{Stanford PULSE Institute}, \orgname{SLAC National Accelerator Laboratory}, \orgaddress{\city{Menlo Park}, \state{California} \postcode{94025}, \country{USA}}}

\affil[3]{\orgdiv{Department of Applied Physics}, \orgname{Stanford University}, \orgaddress{\city{Stanford}, \state{California} \postcode{94305}, \country{USA}}}

\affil[4]{\orgdiv{J.R. Macdonald Laboratory}, \orgname{Kansas State University}, \orgaddress{\city{Manhattan}, \state{Kansas} \postcode{66506}, \country{USA}}}


\abstract{We demonstrate the production of mutually phase-stable attosecond X-ray pulse pairs with tunable relative time delays and phases in a cascaded X-ray free-electron laser. We showcase the method in an experiment at the LCLS-II, in which a shaped electron beam is used in a split undulator configuration to generate the two attosecond pulses. We achieve mutual phase stability by reusing microbunching generated in the first undulator in order to seed the FEL process in the second at a detuned frequency. We measure controllable temporal delays between the two pulses directly in the time domain using angular streaking of photoelectrons, with a step size of 250 attoseconds. We then show that the behavior of the X-ray spectrum is consistent with phase stability between the two pulses, with a relative phase that can be easily tuned using inter-undulator phase shifters. This method is particularly well-suited to few to ten eV energy separations and sub to few femtosecond time delays, which are ideal for experiments in the soft X-ray regime for pushing the limits of our models for molecular dynamics and exerting direct coherent control over quantum systems. }

\maketitle

The motion of bound electrons marks the starting point for all chemical reactions and occurs on timescales of few-femtoseconds to attoseconds.
Probing these extreme timescales requires tools with exquisite temporal resolution. Traditionally, that capability was provided by tabletop light sources leveraging high harmonic generation to generate attosecond pulses in the XUV \cite{paul2001observation,hentschel2001attosecond,ferray1988multiple,krausz2009attosecond,corkum2007attosecond}. 
More recently, the development of attosecond pulses from X-ray free-electron lasers (XFELs) has extended attosecond science into the X-ray regime \cite{huang2017generating,marinelli2017experimental,duris2020tunable,maroju2020attosecond,malyzhenkov2020single,prat2023coherent,franz2024,yan2024terawatt,inoue2025experimental,robles2026broadband,guo2026high}.
Using attosecond X-ray pulses from XFELs, it is possible to drive nonlinear processes \cite{o2020electronic,franz2024,inoue2025experimental}, while taking advantage of the atomic site specificity afforded by X-ray photon energies. 
This has already enabled the study of electronic coherence \cite{li2022attosecond,wang2025probing} and photoemission delays \cite{driver2024attosecond,ji2025attosecond} in molecules, as well as attosecond pump attosecond probe spectroscopy \cite{guo2024,li2024attosecond}.

Most of current understanding comes from a regime of impulsive excitation, i.e. the pump pulses are much shorter than the timescale for the motion of the system~\cite{guo2024, li2022attosecond,wang2025probing,driver2024attosecond,ji2025attosecond}. 
On the other hand, it is natural to attempt to exert control over molecular processes by shaping external fields to either influence the outcome of a molecular reaction~(i.e. product state distributions) or modify the time-dependent properties of the system during the reaction.
The idea of using sculpted laser pulses to drive coherent excitations of electronic wavepackets and overcome limitations imposed by intramolecular vibrational relaxation~(IVR) to influence product state distributions in molecular reactions has been deemed charge-directed reactivity~\cite{remacle1998charge}.
Demonstrating such control schemes would significantly enhance our understanding of electron-nuclear dynamics, as we will have a direct measure of the forces that need to be applied to molecules to drive a certain chemical reaction.
Such control has only preliminary experimental investigation, due largely in part to a lack of control over the XUV and X-ray pulses used for excitation. 
Even if charge directed reactivity is not fully realized in a molecular system, using shaped laser pulses to influence the temporal dynamics still provides an additional degree of freedom for experiments to discern reaction pathways providing a dynamical understanding of molecular processes.  

X-ray pulse shaping in the attosecond and few femtosecond regime has been proposed and demonstrated in several forms before. Most methods have been restricted to amplitude shaping, with limited if any control over the spectral phase. 
The generation of attosecond pulses at all is enabled by electron beam shaping, for example by nonlinear compression \cite{huang2017generating,malyzhenkov2020single,prat2023coherent,yan2024terawatt,prat2025enhanced}, self-modulation \cite{duris2020tunable,macarthur2019phase}, laser modulation \cite{duris2021controllable,zholents2005method}, or seeded microbunching instability techniques \cite{zhang2020experimental,cesar2021electron,li2024beam,robles2026broadband,guo2026high}. 
Independent pairs of short pulses have been generated using double bunch \cite{marinelli2015high} or split undulator techniques \cite{lutman2016fresh}, but these generally produce pulse pairs which are not locked together in phase and with time delays in the many femtosecond range. 
In order to have a stable phase relationship between different colors in a self-amplified spontaneous emission (SASE) FEL, they must somehow be seeded by the same field or electron microbunching. 
This can be accomplished in cascaded schemes in which either microbunching from an earlier stage is reused, or the field generated in an earlier stage seeds lasing in fresh electrons. 
Such cascaded attosecond schemes have been used to generate harmonically related pulse pairs \cite{guo2024}, though with no evidence of mutual phase stability, and to generate high power pulses through FEL superradiance \cite{franz2024,robles2024three}. 
The coherent structure of attosecond pulses can be shaped in cascaded schemes in which the undulator taper is used to control the spectrotemporal shape of the X-rays, a technique which was demonstrated with spectral measurements but without direct evidence in the time domain~\cite{robles2025spectrotemporal}. 

Here, we present evidence for tunable attosecond pulse pairs with mutual phase stability. 
The method we showcase is particularly well-suited to the generation of pairs of attosecond pulses with few to ten eV energy separations and few femtosecond temporal separations. 
This parameter space is ideal for ultrafast coherent control experiments in the soft X-ray regime, since these spectral and temporal gaps are similar to those found between core-to-valence resonant excitations and molecular core-hole lifetimes, respectively, in many molecules containing carbon, oxygen, and nitrogen. 
We directly measure the tunable time delay between the two pulses using an angular streaking technique \cite{itatani2002attosecond,hartmann2018attosecond,li2018characterizing}, and show through comparisons with a start-to-end simulation model that our results are consistent with shot-to-shot phase stability, and that that stable phase can be directly tuned using inter-undulator phase shifters which are commonplace at modern XFEL facilities. 

\section*{Demonstration of attosecond pulse pairs}

\begin{figure*}[htb!]
    \centering
    \includegraphics[width=\linewidth]{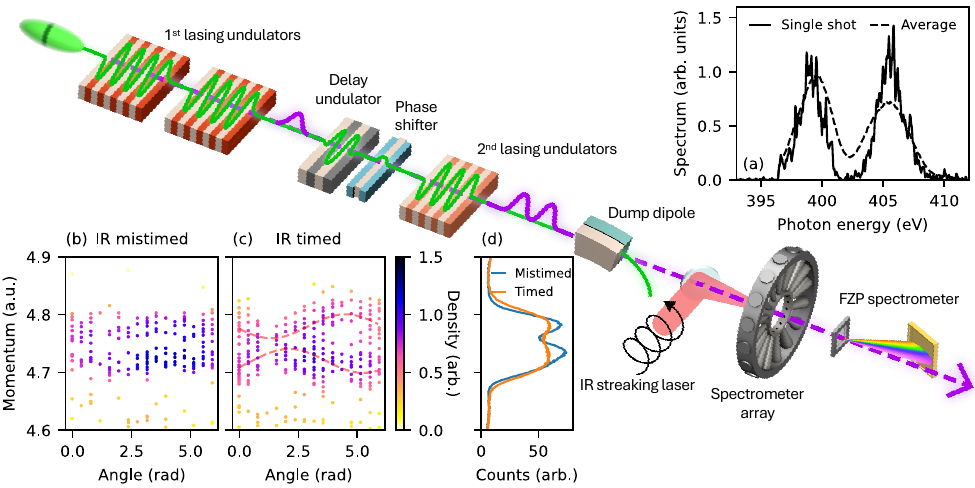}
    \caption{Summary of the experimental setup. A shaped electron bunch emits an isolated attosecond pulse in one set of undulators. Subsequently, the time delay between the beam and radiation can be tuned on the 100 attosecond scale using a delay undulator and on the scale of the radiation period using a phase shifter. Following this, a second pulse is emitted, seeded by the residual microbunching in the beam after generating the first pulse. X-ray properties are characterized in the time domain with angular streaking and spectrally using a Fresnel Zone Plate (FZP). Panel (a) shows single shot and average X-ray spectra. Panels (b) and (c) show single shot measured electron momentum distributions without and with the effect of the streaking laser. The red dashed lines give the best estimate of the streaking kicks received by the two groups of photoelectrons. Panel (d) shows the angle-averaged momentum distribution without and with the streaking laser. }
    \label{fig:schematic}
\end{figure*}

Our experimental setup is shown in Figure~\ref{fig:schematic}. 
The experiment was performed at the Linac Coherent Light Source II~(LCLS-II) using the newly commissioned superconducting accelerator which, at the time of the experiment, generated 3.5~GeV electron beams at a repetition rate of 8~kHz. 
To generate attosecond soft X-ray pulses, we shape the driving electron bunch by controlling the temporal profile of the drive laser for the electron gun \cite{zhang2020experimental,robles2026broadband}. 
This leads to a 10~kA scale sub-femtosecond current spike in the electron bunch through the microbunching instability \cite{borland2002start,saldin2002klystron,saldin2004longitudinal,heifets2002coherent}. 
In one set of undulators, the current spike generates an isolated attosecond pulse~(IAP) with 3~to~4~eV FWHM bandwidth and $\sim15$~$\mu$J pulse energy~(on average). 
On the basis of pulse reconstruction from streaked photoelectrons reported elsewhere \cite{guo2026lclsii}, we estimate the FWHM duration of this pulse to be 500 attoseconds on average. 
At the time of the experiment, the LCLS-II undulator line was split into 3.4 meter long segments, each containing 87 periods with period length $\lambda_u=3.9$ cm. 
Eight undulator segments were used in the first lasing stage.
Following the first stage, the current spike is strongly microbunched around the central frequency of the IAP. 
Furthermore, by intentionally stopping the amplification process slightly short of saturation, the beam quality in the spike may be suitably preserved to enable further emission in an additional undulator stage. 

After this first stage, a single highly detuned undulator section can be used to delay the beam by the total slippage length in the undulator. 
The delay induced in one undulator could be tuned between 120 attoseconds and 2 femtoseconds by changing the undulator gap.
Additionally, a magnetic phase shifter could be used to adjust the timing between the existing microbunching and the IAP on the scale of the radiation period. 
This acts as a fine adjustment of the phase of the microbunching relative to that of the IAP. 
Subsequently, a second, shorter set of only three undulator segments is slightly detuned from the central frequency of the first pulse. 
In this second set of undulators, the pre-bunched beam quickly emits a second pulse comparable in pulse energy to the first despite the shorter undulator length.
This second pulse is expected to have a similar duration to the first, since it derives from the same microbunching structure. 
At this point the total pulse energy doubles to nearly 30 $\mu$J. 
The resonant frequency of the second undulator can be tuned freely, though the amount of microbunching available for seeding at a particular frequency determines the length of undulator needed to emit a strong second pulse. 
Downstream of the undulators, we characterize the total X-ray pulse energy and X-ray spectrum on a single-shot basis using a gas monitor detector and a Fresnel Zone Plate (FZP) spectrometer~\cite{larsen2023compact}, respectively. 
Panel~(a) in Figure~\ref{fig:schematic} shows an example single-shot X-ray spectrum measured in the two-pulse mode alongside the average spectrum. 
The average spectrum is characterized by two roughly Gaussian lobes, separated by 6~eV, each with few~eV FWHM bandwidth, and with similar peak intensities. 

We characterize the X-ray pulse temporal properties using the angular streaking technique \cite{li2018characterizing, hartmann2018attosecond} at the TMO experimental hutch \cite{walter2025time}. 
The X-rays are used to ionize photoelectrons from the 3d orbitals of a gas of Krypton atoms in the presence of a circularly polarized infrared laser, see details in the Methods. 
The photoelectron momenta are characterized shot-by-shot using an array of sixteen time of flight spectrometers. 
The IR laser imparts a time-dependent kick to the ionized electrons which maps the time at which they were ionized into their transverse momentum, conserving the canonical momentum $\vec{p}_c(t)=\vec{p}(t)-e\vec{A}(t)$. 
Figure~\ref{fig:schematic} panels~(b) and (c) show single shot measurements of the photoelectron momentum spectrum when the dressing laser field is intentionally mistimed~(b) and correctly timed~(c), while panel~(d) shows the comparison of the angle-averaged momentum distribution without and with proper timing. 
When the dressing laser is mistimed, the photoelectron spectrum shows two nearly Gaussian features, mimicking the X-ray spectrum shown in panel~(a), with the distinct lower (higher) momentum feature being attributed to the distinct lower (higher) energy part of the X-ray spectrum. 
With the dressing laser properly timed, the two photoemission features receive roughly sinusoidal kicks as a function of the observation direction. 
Panel~(c) of Figure~\ref{fig:schematic} shows red-dashed lines indicating the best fit estimate of the streaking kick received by each photoelectron feature. The fact that both lines appear to be perturbed sinusoidally is evidence that the pulses are significantly shorter than the IR period \cite{li2018characterizing}. The X-ray pulse duration can be estimated more precisely with various algorithms fitting measured dressed-photoelectron features to Strong Field Approximation (SFA) models of the ionization and streaking process~\cite{li2018characterizing,duris2020tunable,franz2024,guo2026lclsii}. 
The fact that the two features appear to be kicked with different phases implies a discrete difference in the arrival times of the two groups of photoelectrons. This gives evidence that the two Gaussian features in the X-ray spectrum are indeed representative of two time-delayed pulses. 

\section*{Stability and tunability of time delay and relative phase}

\begin{figure}[h!]
    \centering
    \includegraphics[width=0.65\linewidth]{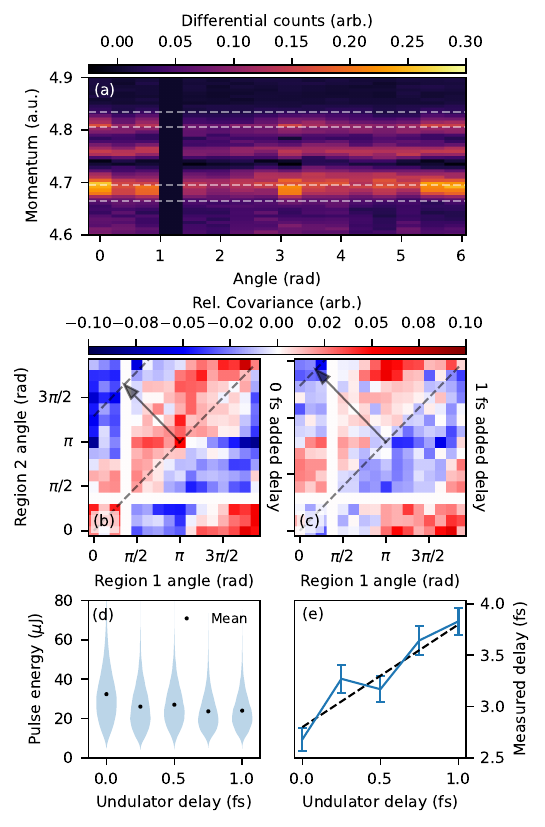}
    \caption{Pulse pair time delay measurement and control. Panel (a) shows the difference in electron momentum distributions with and without the streaking effect. Panels (b) and (c) show angle-angle relative covariance maps for 0 fs and 1 fs delay undulator settings, respectively. Panel (d) shows the total pulse energy as a function of the undulator delay, and panel (e) shows the measured relative delay.}
    \label{fig:delay_scans}
\end{figure}

The time delay between the pulses can be extracted from the relative phase of the sinusoidal kick received by each photoemission feature.
We quantify that phase difference using a covariance method \cite{guo2024,wang2024covariance,driver2024attosecond}. Figure~\ref{fig:delay_scans}(a) shows the electron momentum distribution at each spectrometer as a difference between ``streaked" shots and unstreaked shots, labeled as differential electron counts. 
This is averaged over many independent shots of the machine, which also averages over random arrival times of the IR with respect to the X-rays, hence the sinusoidal behavior is washed out. Regions of positive differential counts are those which electrons have been pushed into by the streaking laser. 
Three positive regions are evident -- the one at the highest momentum almost exclusively contains electrons streaked from the upper photoemission feature. 
The opposite is true for the region of lowest momentum. 
In the middle region, electrons could have reached this momentum from either of the two emission features. The time delay between the emission of the two features can be evaluated by correlating the electron counts below the lower photoemission feature (region 1) with those above the upper photoemission feature (region 2), indicated by the white dashed lines in panel (a). 
If the detection of electrons in the lower region at some spectrometer angle $\theta$ correlates with the detection of electrons in the upper region at a different angle $\phi$, the difference $\phi-\theta$ can be used to extract the time delay between the two photoelectron features in units of the streaking laser phase. 
This technique is described in detail in \cite{wang2024covariance}.

We define the number of electrons found in the spectrometer at angle $\theta$ in the two momentum regions of interest on the $i$th shot of the machine as $X_i(\theta)$ and $Y_i(\theta)$, respectively. 
We evaluate the covariance between the two angles as $C(\theta,\phi)=\langle X_i(\theta)Y_i(\phi)\rangle-\langle X_i(\theta)\rangle\langle Y_i(\phi)\rangle$, where the average is taken over independent shots of the machine. 
We can further improve this using the partial covariance, where fluctuations in shot-to-shot pulse energy are explicitly factored out \cite{wang2024covariance}. 
We then define the relative covariance $\Delta C(\theta,\phi)=C_\text{timed}(\theta,\phi)-C_\text{mistimed}(\theta,\phi)$ -- the difference between the covariance for streaked and unstreaked shots. 
This object can be used to estimate the time delay between the two photoelectron features, as shown in panels (b) and (c). 
A positive value of $\Delta C$ at a pair of angles $\theta$ and $\phi$ implies that the presence of an electron below the lower features at angle $\theta$ is positively correlated with finding an electron above the upper feature at angle $\phi$. 
If the two groups of photoelectrons were emitted with no time delay, we expect perfectly negative correlation for $\theta=\phi$, along the diagonal -- an electron arriving into the lower momentum region should correspond to fewer electrons in the upper region. 
In this case, both panels show an off-diagonal negative line shifted with respect to the main diagonal. 
This shift provides a measure of the time delay between the pulses. 
Two things are worth noting:
First, the covariance is an inherently shot-averaged method, so the fact that these correlations survive is evidence of the stability of the delay between the two pulses. 
Second, when we add a 1 femtosecond delay between the two undulator sections using a delay undulator, the line of negative correlation shifts further from the diagonal, implying that the delay has increased due to the delay undulator, as expected. 

To demonstrate control over the delay, we scanned the strength of the delay undulator to generate delays between $0$ and $1$ femtosecond in $250$ attosecond steps. The FEL dynamics here are in general quite complicated, as several phenomena can happen at once. Inducing a time delay on the electrons also introduces dispersion, which shears the microbunching formed in the first stage. Depending on exactly how far the current spike is from saturation, this can amplify the microbunching through the optical klystron effect \cite{csonka1978enhancement,elleaume1983optical,ding2006optical} or wash it out if the microbunches become overcompressed \cite{guo2024}. This affects the power emitted into the second pulse -- at the same time, the initial IAP is still partially overlapped with the electrons which emitted it, which means that it can undergo amplification or absorption depending on the relative phase and time delay. The relative phase advance was not fixed as we scanned the delay, so the relative phase between the first IAP and the microbunching is not controlled during the scan. Nevertheless, for all delays the two pulses are still present, though the pulse energy shrinks from more than 30 $\mu$J to just over $20$ $\mu$J with a 1 fs delay, as shown in panel (d). Panel (e) shows the delay measured through the relative partial covariance method as a function of the expected slippage in the delay undulator. The result is well-fit by a line with a slope of $1$ and an offset of $2.80\pm0.03$ fs, confirming the simple picture that the delay undulator directly changes the pulse-to-pulse delay despite the complicated underlying FEL dynamics. The nominal delay of 2.8 fs without an explicit delay undulator is consistent with start-to-end simulations shown in the methods.

\begin{figure}[h!]
    \centering
    \includegraphics[width=\linewidth]{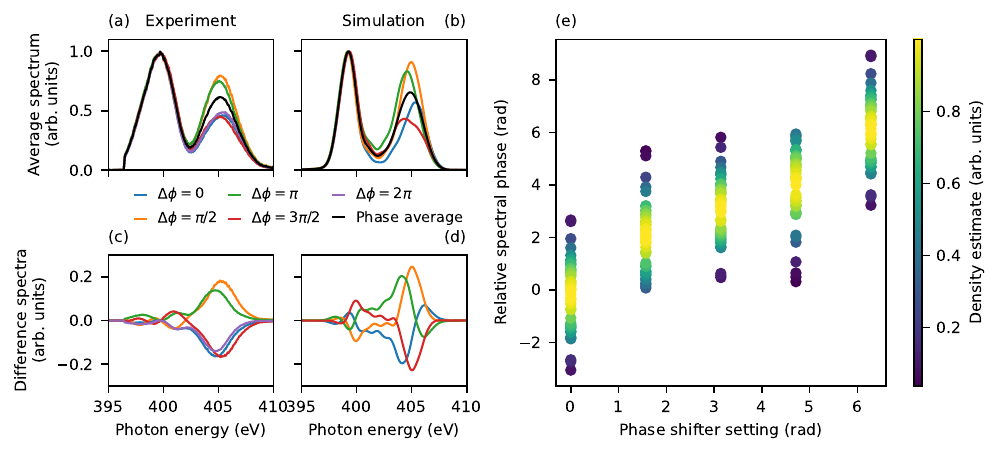}
    \caption{Pulse pair phase stability and control. Panels (a) and (b) show average spectra for different phase shifter settings from experiment and from start-to-end simulations, respectively. Panels (c) and (d) show difference spectra from the same configurations. Panel (e) shows the simulated relative spectral phase difference between the two peaks for 50 different shot noise seeds as a function of the phase shifter setting. }
    \label{fig:phase_scans}
\end{figure}

Because the second pulse is generated by the same microbunching that emitted the first, it should maintain a stable phase relationship with the first pulse which can be tuned using a phase shifter between the two amplification stages. Figure~\ref{fig:phase_scans}(a) shows average spectra measured as a function of a phase shift induced by a phase shifter between the two stages. The most obvious effect is to affect the amplification of the initial pulse, while the second pulse is more weakly affected -- since the second pulse is made entirely in the second stage, it simply inherits the phase of the microbunching at the entrance of the second undulators. The periodic effect of the phase shifter on the amplification of the first pulse is clearest by looking at the difference between each of these spectra and a ``phase averaged" spectrum generated by averaging the results for $\Delta\phi=0$, $\pi/2$, $\pi$, and $3\pi/2$. Those difference spectra are shown in panel (c), where it is clear that phase shifts separated by $\pi$ are roughly mirror images of one another. 

To further illuminate the physics at play, we simulate the experiment with a start-to-end model. We simulate the electron beam dynamics through the accelerator using elegant \cite{borland2000elegant} and the FEL lasing process using GENESIS 1.3 v4 \cite{reiche1999genesis,genesis4github}, the details of the simulation are presented in the methods. Panel (b) shows the average spectra from four simulated phase shifter configurations, each of which has been averaged over 50 simulations seeded by different random initial shot noise and convolved with an estimated FWHM spectral resolution of 0.35 eV \cite{larsen2023compact}. The same qualitative behavior can be seen as in the experiment. Panel (d) shows the difference spectra with respect to the equivalent phase averaged spectrum, and again qualitatively similar behavior is found in which phase shifter configurations separated by $\pi$ mirror one another. In the simulations we can directly compare the relative spectral phase of the field at the two dominant colors. We plot that relative spectral phase as a function of phase shifter setting in panel (e). The points are assigned a color based on a Gaussian estimate of the density at each phase shifter setting. The qualitative behavior we observe is thus consistent with the spectral phase between the two pulses changing by the same step as the phase shifter setting. The actual shot-to-shot phase difference fluctuates due to the initial random shot noise seed. With no filtering, the RMS phase difference fluctuation is $1.31\pm0.04$ rad. This can be shrunk by filtering on the spectral shape. For example, pulses which are fit well by a double Gaussian ($r^2>0.99$, see methods) have RMS phase fluctuation of only $0.81\pm0.03$ rad. This implies that the phase shift between the two pulse is indeed both stable and directly tunable. 

\section*{Conclusions}

In conclusion, we've provided evidence that we can produce mutually phase stable attosecond soft X-ray pulse pairs with direct control over both the time delay and the relative phase. These pulses are ideally suited to coherent control applications in which a pump pulse structured over a few femtoseconds is used to selectively excite different parts of a system at different times and with different relative phases. These pulses can find immediate use in controlling coherent electronic wavepackets in molecules \cite{li2022attosecond,robles:ipac23-tupl094} and extending stimulated X-ray Raman scattering beyond the impulsive regime \cite{o2020electronic}. 

\bibliography{sn-bibliography}

\subsection*{Methods}

\subsubsection*{Accelerator setup}

The LCLS-II is a superconducting accelerator composed of a normal conducting RF photoinjector, three L-band linac sections, two bunch compressors, and a set of third harmonic linearizing cavities. At the time of the experiment, the nominal energy at the end of the linac was 3.5 GeV and the maximum repetition rate was 8 kHz. The photoinjector is driven by a Gaussian laser with a typical FWHM duration of 16 ps. The resulting beam is compressed to 100 femtoseconds by the end of the linac, and must propagate through a 2 km long bypass line to reach the undulators. 

To generate attosecond pulses, we use the photocathode laser shaping technique \cite{zhang2020experimental} to seed the microbunching instability \cite{borland2002start,saldin2002klystron,saldin2004longitudinal,heifets2002coherent}. This involves introducing a second laser on top of the first with a duration of just 3 picoseconds, which introduces a short, small amplitude bump to the current profile of the beam as it exits the cathode. This bump introduces localized collective effects which seed the microbunching instability through the remaining linacs and bunch compressors, and especially through the long bypass line. By the end of the bypass line, the beam can be made to have a localized, mostly single cycle energy modulation that can be compressed once more to create a short, high current spike. 

At the end of the bypass line, the beam is brought to the undulators through a complex series of dispersive sections. A final round of bunch compression can be accomplished in these dispersive sections, where by introducing an orbit bump or tuning quadrupole magnets a variable $R_{56}$ can be generated \cite{nosochkov2021tuning}. We use this variable $R_{56}$ to achieve the ultimate compression of the current spike. 

\begin{figure}[h!]
    \centering
    \includegraphics[width=0.65\linewidth]{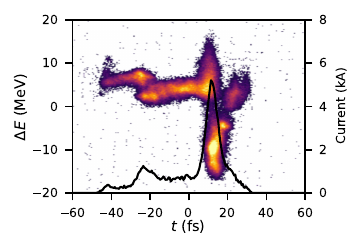}
    \caption{Longitudinal phase space measured at the end of the undulators alongside the current profile. }
    \label{fig:xtcav}
\end{figure}

The electron beam's longitudinal structure is diagnosed using an X-band transverse deflecting cavity (XTCAV) together with a bending magnet oriented perpendicular to the cavity \cite{emma2000transverse,behrens2014few}. The cavity can then streak the time of arrival of different electrons into one transverse degree of freedom while the bend disperses the beam energy onto the other. After these transformations, the beam transverse spot is measured on an Optical Transition Radiation (OTR) screen. Figure~\ref{fig:xtcav} shows the final longitudinal phase space measured after the undulators using the XTCAV system, with an estimated temporal resolution of a few femtoseconds \cite{behrens2014few}. A localized region of high current and high energy spread is evidence of a large current spike in the middle of the beam. Because of the limited resolution, the spike is expected to be much shorter than it appears on the screen, as evidenced in particular by the large blowup of energy spread in both the positive and negative directions attributed to collective effects before and through the undulators. 

As it is due primarily to longitudinal space charge effects associated with current density gradients, the large energy spread around the current spike is expected to be correlated within the sub-femtosecond spike, and compensable by a positive taper of the undulator strength $K$ \cite{ding2009generation}. Theoretically, the appropriate taper rate for a given energy chirp $d\gamma/ds$ is \cite{saldin2006self}
\begin{align}
    \frac{dK}{dz} = \frac{2+K^2}{\gamma K}\frac{\lambda_r}{\lambda_u}\frac{d\gamma}{ds}
\end{align}
in our experiment we used a taper rate corresponding to a chirp of $c\frac{d\gamma}{ds}=10$ MeV/fs, which we found to give us optimal lasing.

\begin{figure}[h!]
    \centering
    \includegraphics[width=\linewidth]{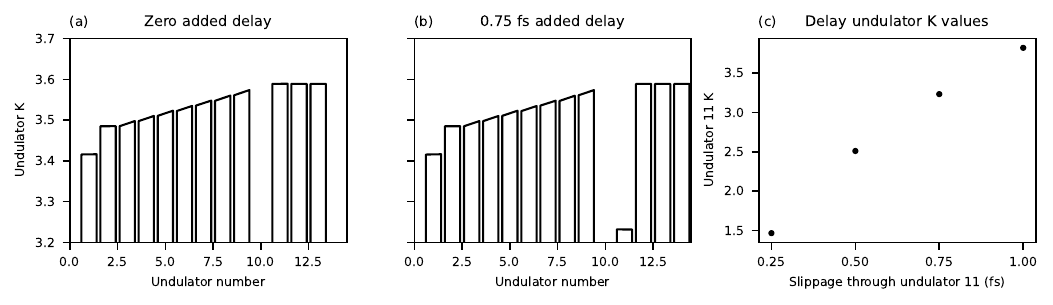}
    \caption{Undulator taper configurations for (a) no added delay and (b) 0.75 fs delay added with undulator 11. Panel (c) shows the $K$ value of undulator 11 to accomplish the delays mentioned in the main text. }
    \label{fig:unds}
\end{figure}

Figure~\ref{fig:unds} shows the undulator taper profile used in the experiments. Panel (a) shows the undulator profile for the case of zero added delay. Undulator 1 is detuned and not significantly contributing to the lasing process. Undulators 2 - 9 make up the first lasing stage in which the first isolated pulse is produced. Undulator 10 is actually a delay chicane installed primarily for self-seeding experiments and was turned off for this study. Undulators 11 - 13 were used to generate the second, detuned pulse for the zero delay case. Panel (b) shows the undulator taper profile for the case in which a delay undulator is used. In this case, the second lasing stage is pushed down to undulators 12 - 14, and undulator 11 is used in  a highly detuned, delay undulator mode. Panel (b) shows in particular the undulator setting for generating 0.75 fs delay in undulator 11, and panel (c) shows the $K$ used for undulator 11 to generate the four delays showcased in the main text. This value is calculated simply from the number of periods in each LCLS-II soft X-ray undulator (87) times the resonant X-ray period in the undulator for a given $K$. The undulator period is $3.9$ cm. The phase shifter (not pictured) used in the phase control experiments is located at the downstream end of undulator 9, right at the end of the first stage. 

\subsubsection*{Angular streaking configuration}

The X-ray pulses are focused by a pair of Kirkpatrick-Baez mirrors to a point just downstream of an angular array of sixteen Time-of-Flight~(ToF) spectrometers \cite{walter2021multi}. This configuration produces an X-ray spot size of roughly 20~$\mu$m diameter, full-width at half maximum~(FWHM) at the collection point of the ToF spectrometers.
The X-ray beam is spatially and temporally overlapped with a circularly-polarized infrared laser~(2.04~$\mu$m).
Krypton gas is introduced into the chamber by an effusive needle close to the overlap region.
The X-ray photon energy was tuned to $\sim400$~eV, and the ToF retardation was optimized for resolving the electron emission leading to the $3d_{5/2}^{-1}$ and $3d_{3/2}^{-1}$ final states of the krypton ion (with a binding energy of $93.8$~eV and $95.0$~eV respectively).
These electrons will then have a kinetic energy of $\sim306.2$ and $\sim305$~eV, respectively. 

\subsubsection*{Start-to-end simulations}

\begin{figure}[h!]
    \centering
    \includegraphics[width=0.75\linewidth]{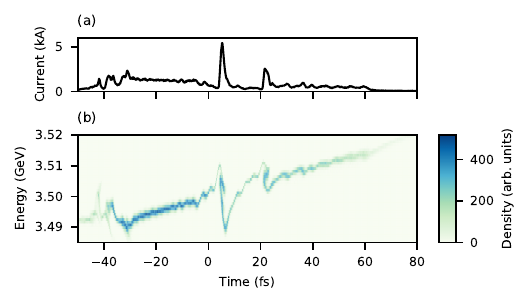}
    \caption{Summary of the longitudinal structure of the electron beam at the undulator entrance in the start-to-end simulations. Panel (b) shows the longitudinal phase space while panel (a) shows the current profile. }
    \label{fig:sim_lps}
\end{figure}

The start-to-end model of the LCLS-II starts with a model of the photoinjector built in the ASTRA code \cite{flottmann2003recent}. After the beam becomes ultra-relativistic, it is passed to the code elegant \cite{borland2000elegant} for propagation to the entrance of the undulators, after which FEL simulations are performed using GENESIS 1.3 v4 \cite{reiche1999genesis,genesis4github}. Figure~\ref{fig:sim_lps} shows the longitudinal structure of the beam at the entrance to the undulators. Panel (a) shows the current profile and panel (b) shows the longitudinal phase space. A current spike with peak current above 5 kA is present around the 5 fs mark, on top of a lower current smooth background with some additional structures. The spike at 5 fs is the one we focus on in the simulations. 

\begin{figure}[h!]
    \centering
    \includegraphics[width=0.7\linewidth]{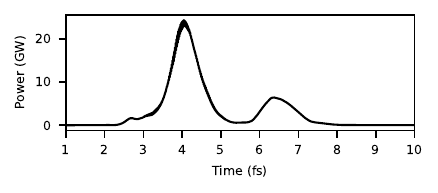}
    \caption{Simulated power profile at the undulator end for no added delay.}
    \label{fig:sim_powers}
\end{figure}

Figure~\ref{fig:sim_powers} shows a single shot power profile simulated in GENESIS 1.3 v4 after passage through the first set of linearly tapered undulators and the second set of three untapered undulators, as in the experiment. Two pulses are evident with a time delay between them of around 2.5 femtoseconds, similar to the ``zero added delay" pulse to pulse delay found in the experiment. We found that the simulated spectra shown in Fig.~\ref{fig:phase_scans} match the qualitative behavior of the experimental ones only if the simulated phase shifter setting was $\pi/2$ different from the nominal experimental value. This is likely due to imprecision in the knowledge of the absolute setting of the phase shifter for our particular lasing wavelength, as the phase shifters are designed for normal FEL operations where the undulator $K$ values vary more smoothly along the undulator line. 

\subsubsection*{Coefficient of determination}

For a sequence of data points $y_i$ fit by model values $x_i$, the coefficient of determination, or $r^2$, is defined as 
\begin{align}
    r^2 = 1 - \frac{\sum_i(y_i-x_i)^2}{\sum_i(y_i-\bar{y})^2}
\end{align}
where $\bar{y}$ is the average value of the $y_i$.

\subsection*{Acknowledgements}
Use of the Linac Coherent Light Source (LCLS), SLAC
National Accelerator Laboratory, is supported by the
U.S. Department of Energy, Office of Science, Office of
Basic Energy Sciences under Contract No. DE-AC02-
76SF00515
This work was primarily supported by the U.S. Department of Energy, Office of Science, Office of Basic Energy Sciences under
Contract No. DE-AC02-76SF00515, and by the U.S. Department of Energy, Office of Science, Office of Basic Energy Sciences Accelerator and Detector Research Program. 
R. R. R. acknowledges the support of the Robert H. Siemann Fellowship. 
JPC, TD, and JW were supported by the US DOE, Office of Science, Office of Basic Energy Sciences (BES), Chemical Sciences, Geosciences, and Biosciences Division (CSGB). 

\end{document}